\documentclass{JHEP3} % 10pt is ignored!

\usepackage{epsfig,multicol,bbm}
\usepackage{cite}

\newcommand\fverb{\setbox\fverbbox=\hbox\bgroup\verb}
\newcommand\fverbdo{\egroup\medskip\noindent%
            \fbox{\unhbox\fverbbox}\ }
\newcommand\fverbit{\egroup\item[\fbox{\unhbox\fverbbox}]}
\newbox\fverbbox
\newcommand{\nb}{\nonumber}

\newcommand{\qqJZ}{$q\bar q \to J/\psi +Z^0~$}
\newcommand{\ggJZ}{$gg \to J/\psi +Z^0~$}
\newcommand{\ppJZ}{$pp \to J/\psi +Z^0+X~$}

\title{  QCD corrections to $J/\psi$ plus $Z^0$-boson production at the LHC }

\author{Song Mao$^a$, Ma Wen-Gan$^a$, Li Gang$^b$, Zhang Ren-You$^a$,  and Guo Lei$^a$\\
$^a$Department of Modern Physics, University of Science and Technology of China,
Hefei, Anhui 230026, P.R.China \\
$^b$School of Physics and Material Science, Anhui University, Hefei, Anhui 230039, P.R.China\\
E-mail: \email{songmao@mail.ustc.edu.cn }, \email{mawg@ustc.edu.cn},
\email{lig2008@mail.ustc.edu.cn}, \email{zhangry@ustc.edu.cn},
\email{guolei@mail.ustc.edu.cn}}

\vskip 15mm \abstract{ The $J/\psi+Z^0$ associated production at the
LHC is an important process in investigating the color-octet
mechanism of non-relativistic QCD in describing the processes
involving heavy quarkonium. We calculate the next-to-leading order
(NLO) QCD corrections to the $J/\psi +Z^0$ associated production at
the LHC within the factorization formalism of nonrelativistic QCD,
and provide the theoretical predictions for the distribution of the
$J/\psi$ transverse momentum. Our results show that the differential
cross section at the leading-order is significantly enhanced by the
NLO QCD corrections. We conclude that the LHC has the potential to
verify the color-octet mechanism by measuring the $J/\psi+Z^0$
production events.  }


\vfill \eject

\begin{document}

\par
\section{Introduction}
\par
Since the discovery of the first charm-anticharm bound state
$J/\psi$, the study of quarkonium is a very interesting topic in
both a theoretical and an experimental point of view, which provide
a good place to probe both perturbative and non-perturbative aspects
of QCD dynamics.

\par
In the early days of quarkonium physics, the production and decay of
a quarkonium was described by the color-singlet mechanism (CSM)
\cite{Einhorn:1975ua,Einhorn:1976,chang:1980,derger:1981,baier:1982}
with the color-singlet $Q\bar{Q}$ pair having the appropriate spin,
angular-momentum and charge conjugation quantum numbers, and it was
assumed that the decay and production of a quarkonium can be
factorized into a short distance part, which can be calculated in a
perturbative series of the running coupling constant $\alpha_s(M)$,
and a long distance part, which relies on the nonperturbative
dynamics of the bound state. However, the CSM has encounter many
difficulties in various theoretical
\cite{Barbieri:1976fp,Barbieri:1980,Barbieri:1981,Barbieri:1979,hagiwara:1981,mackenzie}
and experimental aspects \cite{Abe:1992ww,Abe:1993ww}, such as the
appearance of a logarithmic infrared divergence in the case of NLO
$P$-wave decays into light hadrons and the huge discrepancy of the
high-$p_t$ $J/\psi$ production between the theoretical prediction
and the experimental measurement at the Tevatron.

\par
In 1995, the nonrelativistic QCD (NRQCD) \cite{bbl}, proposed by
Bodwin, Braaten and Lepage (BBL), provides a rigorous theoretical
framework for the description of heavy-quarkonium production and
decay. In the NRQCD, the idea of perturbative factorization is
retained, the process of production and decay of heavy quarkonium is
separated into two parts: short distance part, which allows the
intermediate $Q\bar{Q}$ pair with quantum numbers different from
those of the physical quarkonium state, and the long distance matrix
elements (LDMEs), which can be extracted from experiments. The
relative importance of the LDMEs can be estimated by means of the
velocity scaling rules \cite{Lepage:1992}. If one only retains the
lowest order in $v$, the description of $S$-wave quarkonia
production or annihilation reduces to the CSM. In the case of
$P$-waves, infrared singularities which appear in some of the short
distance coefficients can be absorbed into the long distance part of
color-octet $S$-wave states. Including the contribution of
color-octet states, the large discrepancies between the experimental
data of $J/\psi$ production at the Tevatron and the theoretical
predictions based on CSM successfully reconciled
\cite{jpsi:com-a,jpsi:com-b,jpsi:com-c,jpsi:com-d}.

\par
Recently, substantial progress has been achieved in the calculation
of high order QCD corrections to $J/\psi$ hadroproduction in order
to clarify the validity and limitation of the NRQCD formulism. The
DELPHI data more favor the NRQCD color-octet mechanism (COM)
predictions for the $\gamma\gamma \to J/\psi+ X$ process
\cite{DELPH1,DELPH2}. Similarly the recent experimental data on the
$J/\psi$ photoproduction of H1 \cite{H1} are fairly well described
by the complete NLO NRQCD corrections \cite{kniehl1}, and give
strong support to the existence of the COM. However, At B-factories,
a series of processes was calculated up to the QCD NLO corrections
in the CSM
\cite{zhang:092001,zhang:092003,zhang:054006,zhang:162002,wang:054028,wang:181803,
wang:162003,sang:034028,chao:114014,zhang:034015,wjx1}. Together
with the relativistic correction \cite{RC-a,RC-b,RC-c,RC-d,RC-e}, it
seems that most experimental data could be understood. Additionally,
the $J/\psi$ polarization in hadroproduction at the Tevatron
\cite{nnrtev-a,nnrtev-b} and photoproduction at the HERA
\cite{nnrHERA} also conflict with the NRQCD predictions. Therefore,
the $J/\psi$ production mechanism is still a big challenge. The
further tests for the CSM and COM in NRQCD are still needed on the
production of heavy quarkonium.

\par
In order to investigate the effects of the COM in heavy quarkonium
production, it is an urgent task to study the processes which
heavily depend on the production mechanism. The $J/\psi$ production
associated with a gauge boson at the LHC, is a suitable process for
studying COM. In high energy collider experiments, $W^\pm$, $Z^0$
and $J/\psi$ can be identified by using their purely leptonic decays
\cite{bager:111, braaten:091501}, which are particularly useful in
hadron colliders, because they provide an enormous suppression of
the background. In reference \cite{B.A.Kniehl-lo}, the authors give
a theoretical prediction at the LO for associated production of
heavy quarkonim and electroweak bosons at hadron colliders. The
numerical results show that for $J/\psi+\gamma$ associated
production the CSM provides the main contribution, but for $J/\psi$
+ $W$ and $J/\psi$ + $Z^0$ associated production, the COM
contribution to the cross section is dominative. The NLO QCD
corrections to $pp \to J/\psi + \gamma$ production in the CSM were
provided in \cite{wangphoto}, and the results show that the cross
section in large $p_T$ region of $J/\psi$ can be enhanced by two
orders in magnitude. Recently, the complete NLO QCD corrections to
$J/\psi + W$ were calculated at the LHC in nonrelativistic quantum
chromodynamics \cite{liw}. There only the color-octet contributes to
the process $pp \to J/\psi + W$ at both LO and QCD NLO. The
numerical results show that the differential cross section at the LO
is significantly enhanced by the NLO QCD corrections. In this paper
we calculate the NLO QCD corrections to the associated $J/\psi$
production with a $Z^0$ gauge boson in the NRQCD at the LHC, and
present the theoretical predictions for the $p_T$ distribution of
$J/\psi$.

\par
This paper is structured as follows: In Sec. II we give the
calculation description of the LO cross section for the \ppJZ
process, and the calculations of the NLO QCD corrections are
provided in Sec. III. In Sec. IV we present some numerical results
and discussion, and finally a short summary is given.

\vskip 5mm
\section{LO calculations for the \ppJZ process}
\par
\begin{figure}[htbp]
\vspace*{-0.3cm} \centering
\includegraphics[scale=0.8]{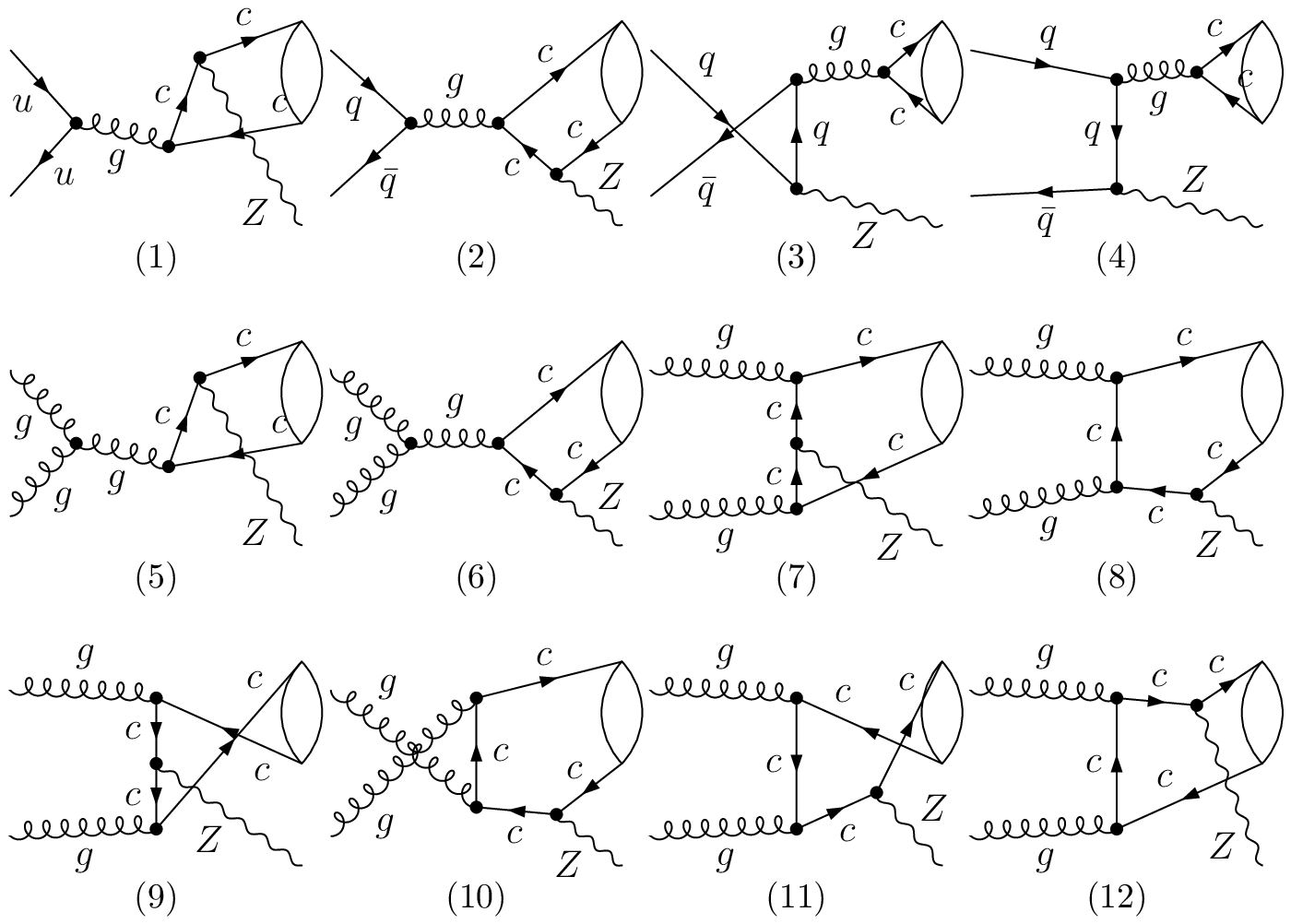}
\vspace*{-0.3cm} \centering \caption{\label{fig1} The tree-level
Feynman diagrams for the partonic processes \qqJZ(1-4) and
\ggJZ(5-12).}
\end{figure}

At the LO, there involves two types of partonic processes, which
contribute to the \ppJZ process:
\begin{equation}
\label{process} q\bar{q} \to c\bar{c}[n] + Z^0,  n = {}^1S_0^{(8)},
{}^3S_1^{(8)}, {}^3P_J^{(8)},
\end{equation}
\begin{equation}
\label{process-1} gg \to c\bar{c}[n] + Z^0,  n =
{}^3S_1^{(1)},{}^1S_0^{(8)}, {}^3S_1^{(8)}, {}^3P_J^{(8)},
\end{equation}
where $q$ represents all possible light-quarks(u, d and s). The
tree-level Feynman diagrams are shown in Fig.\ref{fig1}. The cross
section for the production of a $c\bar{c}$ pair in a Fock state $n$,
$\hat{\sigma}[ij\to c\bar{c}[n]+Z^0]$, is calculated from the
amplitudes which are obtained by applying certain projectors onto
the usual QCD amplitudes for open $c\bar{c}$ production. In the
notations of Ref.\cite{p2}:
\begin{eqnarray}
{\cal A}_{c\bar{c} [{}^1S_0^{(8)} ]} = {\rm Tr} \Big[ {\cal C}_8
\Pi_0 {\cal A} \Big]_{q=0}, \nonumber
\end{eqnarray}
\begin{eqnarray}
{\cal A}_{c\bar{c} [ ^3S_1^{(1/8)} ]} = {\cal E}_{\alpha} {\rm Tr}
\Big[ {\cal C}_{1/8} \Pi_{1}^{\alpha} {\cal A} \Big]_{q=0},
\nonumber
\end{eqnarray}
\begin{eqnarray}
{\cal A}_{c\bar{c} [ ^3P_J^{(8)} ]} = {\cal E}_{\alpha \beta}^{(J)}
\frac{d}{dq_{\beta}} {\rm Tr} \Big[ {\cal C}_8 \Pi_1^{\alpha} {\cal
A} \Big]_{q=0}, \nonumber
\end{eqnarray}
where ${\cal A}$ denotes the QCD amplitude with amputated charm
spinors, the lower index q represents the momentum of the heavy
quark in the $Q\bar{Q}$ rest frame. $\Pi_{0/1}$ are spin projectors
onto the spin singlet and spin triplet states. ${\cal C}_{1/8}$ are
color projectors onto the color singlet and color octet states, and
${\cal E}_{\alpha}$ and ${\cal E}_{\alpha \beta}$ represent the
polarization vector and tensor of the $c\bar{c}$ states,
respectively.
%%figure%%
%%%%figure%%%%

\par
Then the LO short distance cross section for the partonic process
$i(p_1)j(p_2)\to c\bar{c}[n](p_3)+Z^0(p_4)$ is obtained by using the
following formula:
\begin{eqnarray}
\hat{\sigma}\left(ij\to c\bar{c}[n]+Z^0\right)=\frac{1}{16\pi
\hat{s}^2 } \int_{\hat{t}_{min}}^{\hat{t}_{max}}
d\hat{t}\overline{\sum}|{\cal A}_{LO}|^2,~~(ij=u\bar u,d\bar d,s\bar
s,gg).
\end{eqnarray}
The summation is taken over the spins and colors of initial and
final states, and the bar over the summation denotes averaging over
the spins and colors of initial partons. The Mandelstam variables
are defined as:
$\hat{s}=(p_1+p_2)^2,\hat{t}=(p_1-p_3)^2,\hat{u}=(p_1-p_4)^2$.

\par
The LO cross section for the \ppJZ process is expressed as
\begin{eqnarray}
\sigma\left(pp \to
J/\psi+Z^0+X\right)&=&\sum_{i,j,n}\frac{1}{1+\delta_{ij}} \int
dx_1dx_2\hat{\sigma}\left(ij\to c\bar{c}[n]+Z^0\right)\frac{<{\cal
O}_{n}^{J/\psi}>}{N_{col}N_{pol}} \nb \\
&&\times \left[G_{i/A}(x_1,\mu_f)G_{j/B}(x_2,\mu_f)+(A
\leftrightarrow B )\right],
\end{eqnarray}
where $G_{i,j/A,B}$ are the parton distribution functions (PDFs). A
and B refer to protons at the LHC, $N_{col}$ and $N_{pol}$ refer to
the numbers of colors and polarization states separately \cite{p2}.
The hadronic matrix elements $<{\cal O}_{n}^{J/\psi}>$ are related
to the hadronization from the states $c\bar{c}[n]$ into $J/\psi$
which are fully governed by the nonperturbative QCD effects. The
indices $i,j$ run over all the partonic species and $n$ denotes the
specific  color, spin and angular momentum $c\bar{c}$-state.

\par
Our numerical calculation shows that the main contributions come
from the $gg \to c\bar{c}[^3S_1^{(1)}] + Z^0$ and $q\bar{q} \to
c\bar{c}[^3S_1^{(8)}] + Z^0$ partonic processes. The contributions
from other partonic processes are relative small. Therefore, we
consider only the NLO QCD corrections to the $gg \to
c\bar{c}[^3S_1^{(1)}] + Z^0$ and $q\bar{q} \to c\bar{c}[^3S_1^{(8)}]
+ Z^0$ partonic processes in the following NLO calculation.

\vskip 5mm
\section{NLO QCD corrections to the \ppJZ process}
\par
In our NLO QCD calculations we consider the following contribution
components for the \ppJZ process:

\par
$\blacktriangleright$ the virtual corrections to the partonic
process $g(p_1)g(p_2) \to c\bar{c}[^3S_1^{(1)}](p_3) + Z^0(p_4)$ and
$q(p_1)\bar{q}(p_2) \to c\bar{c}[^3S_1^{(8)}](p_4) + Z^0(p_4)$.

\par
$\blacktriangleright$ the real gluon emission partonic processes
$g(p_1)g(p_2) \to c\bar{c}[^3S_1^{(1)}](p_3) + Z^0(p_4) + g(p_5)$,
$q(p_1)\bar{q}(p_2) \to c\bar{c}[^3S_1^{(1)}](p_3) + Z^0(p_4) +
g(p_5)$ and $q(p_1)\bar{q}(p_2) \to c\bar{c}[^3S_1^{(8)}](p_3) +
Z^0(p_4) + g(p_5)$.

\par
$\blacktriangleright$ the real light-(anti)quark emission partonic
processes  $q(\bar{q})(p_1)g(p_2) \to c\bar{c}[^3S_1^{(1)}](p_3) +
Z^0(p_4) + q(\bar{q})(p_5)$ and $q(\bar{q})(p_1)g(p_2) \to
c\bar{c}[^3S_1^{(8)}](p_3) + Z^0(p_4) + q(\bar{q})(p_5)$.

\par
$\blacktriangleright$ the collinear counterterms of the PDFs.

\par
\subsection{virtual corrections}
\par
\begin{figure}[htbp]
\vspace*{-0.3cm} \centering
\includegraphics[scale=0.8]{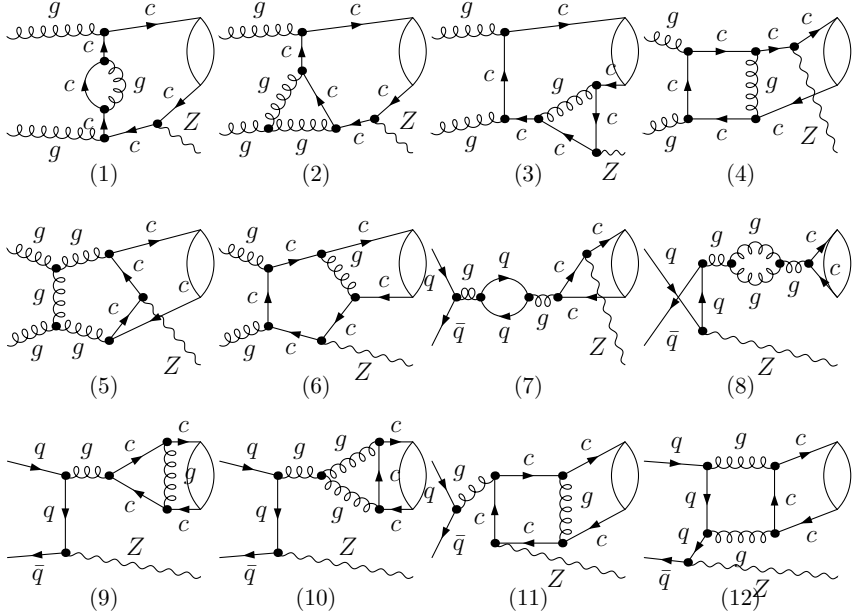}
\vspace*{-0.3cm} \centering \caption{\label{fig2} Some
representative QCD one-loop Feynman diagrams for the partonic
processes $gg \to c\bar{c}[^3S_1^{(1)}] + Z^0$ (1-6) and $q\bar{q}
\to c\bar{c}[^3S_1^{(8)}] + Z^0$ (7-12).}
\end{figure}

The QCD ${\cal O} ({\alpha_s})$ virtual corrections come from the
one-loop diagrams including self-energy, vertex, box, pentagon and
counterterm diagrams. Some representative one-loop Feynman diagrams
are displayed in Fig.\ref{fig2}. We use FeynArts to generate the
diagrams for all related subprocesses \cite{hahn1}. Then we use our
in-house Mathematica program to simplify and square the amplitudes.
The phase space integration is implemented by applying FormCalc
programs \cite{hahn2}. There are UV, IR and Coulomb singularities in
the the virtual corrections. We adopt the dimensional regularization
(DR) scheme to regularize the UV and IR divergences, and the
modified minimal subtraction $\overline{{\rm MS}}$ and on-mass-shell
schemes to renormalize the strong coupling constant and the quark
wave functions, respectively. We adopt the definitions of one-loop
integral functions in Ref \cite{Passarino,denner2}, and use the
Passarino-Veltman reduction formulas to reduce the tenser integrals
to scalar integrals.
%%figure%%
%%%%figure%%%%

\par
We use the expressions in Ref.\cite{IRDV} to deal with the IR
singularities in Feynman integrals, and apply the expressions in
Refs.\cite{OneTwoThree,Four,Five} to implement the numerical
evaluations for the finite parts of N-point scale integrals. In the
virtual correction calculations, we find that Fig.\ref{fig2}(6,9,11)
contain Coulomb singularities, which are regularized by a small
relative velocity $v$ between $c$ and $\bar{c}$ \cite{coulomb}, and
are canceled by those stemming from the NLO QCD correction to the
operator $<{\cal O}_{n}^{J/\psi}>$.

\boldmath
\subsection{Real gluon/light-(anti)quark emission corrections}
\unboldmath
%%figure%%
\begin{figure}[htbp]
\vspace*{-0.3cm} \centering
\includegraphics[scale=0.8]{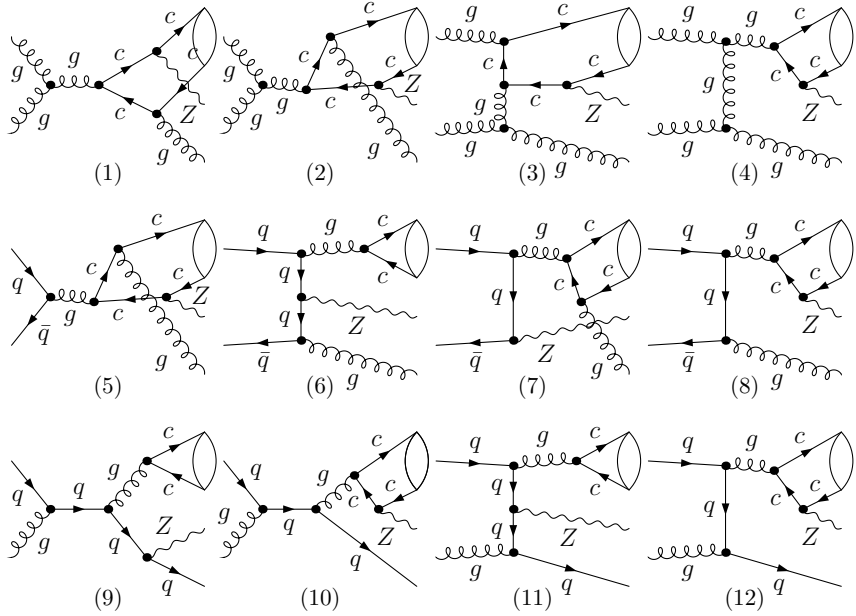}
\vspace*{-0.3cm} \centering \caption{\label{fig3} Some
representative Feynman diagrams for real gluon/light-(anti)quark
emission partonic processes.}
\end{figure}
%%%%figure%%%%

\par
For the real emission partonic processes, there are IR singularities
in the phase space integration. We use the two cutoff phase space
slicing method (TCPSS) to perform the integration over the phase
space of these real emission partonic processes \cite{TCPSS}. In our
calculations, the real gluon emission correction contains both soft
and collinear IR singularities, which are isolated in soft gluon
region($E_5 \leq \delta_s \sqrt{\hat s} /2$) and hard gluon
region($E_5 > \delta_s \sqrt{\hat s} /2$) respectively. It is easy
to find that soft singularities caused by the diagrams emitting soft
gluon from the charm-quark and anti-charm-quark respectively in
$J/\psi$ are canceled with each other. The hard gluon region is
divided into the hard collinear region (HC),i.e., $-\hat{t}_{15}$(or
$-\hat{t}_{25}$)$<\delta_c \hat{s}$, and the hard noncollinear
region ($\overline{HC}$), i.e., $-\hat{t}_{15}$(and
$-\hat{t}_{25}$)$>\delta_c \hat{s}$ where
$\hat{t}_{15}\equiv(p_1-p_5)^2$ and $\hat{t}_{25}\equiv(p_2-p_5)^2$.
Each of the real light-(anti)quark emission processes contains only
collinear IR singularity, and can be dealt with in the hard
collinear region too. In the $\overline{HC}$ region, the real
emission corrections are finite and can be calculated numerically
with general Monte Carlo method. After summing the virtual and real
gluon/light-(anti)quark radiation corrections, the remained
collinear divergence can be cancelled by the collinear counterterms
of the PDFs. By adding all the contributions together, a finite
total cross section is obtained.

\par
The cross sections for the partonic processes $gg \to
c\bar{c}[^3S_1^{(1)}]+Z^0+g$ and $q\bar{q}\to
c\bar{c}[^3S_1^{(8)}]+Z^0+g$ in the soft region can be expressed as
\begin{eqnarray}
\label{soft cross section} \hat{\sigma}^S\left(gg\to
c\bar{c}[^3S_1^{(1)}]+Z^0+g\right) &=&-\frac{\alpha_s}{2\pi}\cdot
(-3) \cdot g(p_1,p_2) \hat{\sigma}_{LO}[^3S_1^{(1)}],
\end{eqnarray}
\begin{eqnarray}
\label{soft cross section-1} \hat{\sigma}^S\left(q\bar{q}\to
c\bar{c}[^3S_1^{(8)}]+Z^0+g\right) &=&-\frac{\alpha_s}{2\pi}\left
[\frac{1}{6}g(p_1,p_2)-\frac{7}{6}\left(g(p_1,p_c)+g(p_2,p_{\bar{c}})\right)\right.
\nb\\
&&\left. -\frac{1}{3}(g(p_1,p_{\bar{c}})+g(p_2,p_c))\right
]\hat{\sigma}_{LO}[^3S_1^{(8)}],
\end{eqnarray}
where $g(p_i,p_j)$ is soft integral function defined as
\cite{c2-been-a,c2-been-b,c2-willy,c2-hs}
\begin{eqnarray}
g(p_i,p_j)=\frac{(2\pi\mu)^{2\epsilon}}{2\pi}\int_{E_5\leq\delta_s\sqrt{\hat{s}}/2}
\frac{d^{D-1}p_5}{E_5}\left[\frac{2(p_ip_j)}{(p_ip_5)(p_jp_5)}-\frac{p^2_i}
{(p_ip_5)^2}-\frac{p^2_j}{(p_jp_5)^2}\right],
\end{eqnarray}
while the cross section of partonic process $q\bar{q} \to
c\bar{c}[^3S_1^{(1)}] + Z^0 + g$ is finite and can be evaluated
directly in four-dimension using Monte Carlo method.

\par
The collinear counterterms of the PDFs, $\delta G_{i/P}(x,\mu_f)~
(P=p~;~i=g, u, \bar{u}, d, \bar{d}, s, \bar{s})$ contains two parts.
One is the collinear gluon emission part $\delta
G_{i/P}^{(gluon)}(x,\mu_f)$, another is the collinear
light-(anti)quark emission part $\delta G_{i/P}^{(quark)}(x,\mu_f)$.
Their analytical expressions are presented as follows.
\begin{eqnarray}\label{PDFcounterterm1}
&& \delta G_{q(g)/P}(x,\mu_f) = \delta G_{q(g)/P}^{(gluon)}(x,\mu_f)
                            +\delta G_{q(g)/P}^{(quark)}(x,\mu_f),
                            ~~(q = u, \bar{u}, d, \bar{d}, s, \bar{s}),
\end{eqnarray}
where
\begin{eqnarray}\label{PDFcounterterm2}
&& \delta G_{q(g)/P}^{(gluon)}(x,\mu_f) =
   \frac{1}{\epsilon} \left[
                      \frac{\alpha_s}{2 \pi}
                      \frac{\Gamma(1 - \epsilon)}{\Gamma(1 - 2 \epsilon)}
                      \left( \frac{4 \pi \mu_r^2}{\mu_f^2} \right)^{\epsilon}
                      \right]
   \int_z^1 \frac{dz}{z} P_{qq(gg)}(z) G_{q(g)/P}(x/z,\mu_f), \nonumber \\
&& \delta G_{q/P}^{(quark)}(x,\mu_f) =
   \frac{1}{\epsilon} \left[
                      \frac{\alpha_s}{2 \pi}
                      \frac{\Gamma(1 - \epsilon)}{\Gamma(1 - 2 \epsilon)}
                      \left( \frac{4 \pi \mu_r^2}{\mu_f^2} \right)^{\epsilon}
                      \right]
   \int_z^1 \frac{dz}{z} P_{qg}(z) G_{g/P}(x/z,\mu_f),  \nonumber \\
&& \delta G_{g/P}^{(quark)}(x,\mu_f) =
   \frac{1}{\epsilon} \left[
                      \frac{\alpha_s}{2 \pi}
                      \frac{\Gamma(1 - \epsilon)}{\Gamma(1 - 2 \epsilon)}
                      \left( \frac{4 \pi \mu_r^2}{\mu_f^2} \right)^{\epsilon}
                      \right]
   \sum_{q=u,\bar{u}}^{d,\bar{d}, s, \bar {s}}
   \int_z^1 \frac{dz}{z} P_{gq}(z) G_{q/P}(x/z,\mu_f).
\end{eqnarray}

\par
The soft and collinear IR singularities from virtual corrections can
be canceled partially by adding with the contributions of the real
gluon/light-(anti)quark emission processes and the gluon emission
part of the PDF counterterms $\delta G_{q(g)/P}^{(gluon)}$. And the
remaining collinear IR singularity can be canceled by the
contributions of the collinear light-(anti)quark emission part of
the PDF counterterms $\delta G_{q(g)/P}^{(quark)}$ exactly. All
these cancelations are verified numerically in our numerical
calculations. The explicit expressions for the splitting functions
$P_{ij}(z),~(ij=qq,qg,gq,gg)$ can be found in Ref.\cite{TCPSS}.

\boldmath
\subsection{NLO QCD corrected cross sections}\unboldmath
\par
The NLO QCD corrected hadronic cross section for $J/\psi+Z^0$
associated production at the LHC can be written as
\begin{eqnarray}
\sigma^{QCD} &=& \sigma^{0} + \Delta \sigma^{QCD} = \sigma^{0} +
\Delta \sigma^{gg}_{{}^3S_1^{(1)}} + \Delta
\sigma^{q\bar{q}}_{{}^3S_1^{(1)}} + \Delta
\sigma^{q\bar{q}}_{{}^3S_1^{(8)}}
\end{eqnarray}

\par
The NLO QCD correction $\Delta \sigma^{QCD}$ contains three
components: (1) $\Delta \sigma^{gg}_{{}^3S_1^{(1)}}$, the NLO QCD
correction to the subprocess $gg \to c\bar{c}[{}^3S_1^{(1)}] + Z^0$,
(2) $\Delta \sigma^{q\bar{q}}_{{}^3S_1^{(1)}}$, the NLO QCD
correction to the subprocess $q\bar{q} \to c\bar{c}[{}^3S_1^{(1)}] +
Z^0$, (3) $\Delta \sigma^{q\bar{q}}_{{}^3S_1^{(8)}}$, the NLO QCD
correction to the subprocess $q\bar{q}\to c\bar{c}[{}^3S_1^{(8)}] +
Z^0$. Due to the absence of the LO cross section for the $q\bar{q}
\to c\bar{c}[{}^3S_1^{(1)}] + Z^0$, the $\Delta
\sigma^{q\bar{q}}_{{}^3S_1^{(8)}}$ contains only the contribution of
subprocess $q\bar{q} \to c\bar{c}[{}^3S_1^{(1)}] + Z^0 + g$ .

\par
The $\Delta \sigma^{gg}_{{}^3S_1^{(1)}}$ and $\Delta
\sigma^{q\bar{q}}_{{}^3S_1^{(8)}}$ can be expressed as:
\begin{eqnarray}
\Delta \sigma^{gg}_{{}^3S_1^{(1)}}= \sigma_{V}^{(gg)} +
\sigma^{(gg)g}_{R}+ \sigma^{(gg)q}_{R}+ \sigma^{(gg)\bar{q}}_{R},
\end{eqnarray}
\begin{eqnarray}
\Delta \sigma^{q\bar{q}}_{{}^3S_1^{(8)}}= \sigma_{V}^{(q\bar{q})} +
\sigma^{(q\bar{q})g}_{R}+ \sigma^{(q\bar{q})q}_{R}+
\sigma^{(q\bar{q})\bar{q}}_{R},
\end{eqnarray}
where $\sigma^{(ij)}_{V}$ , $\sigma^{(ij)g}_{R}$, $\sigma^{(ij)
q}_{R}$ and $\sigma^{(ij)\bar{q}}_{R}$ ($ij=gg,q\bar q$) represent
the virtual, the real gluon emission, the real light-quark and
light-antiquark emission corrections to the cross section,
respectively.

\vskip 10mm
\section{Numerical results and discussion}
\par
As a check of the correctness of our calculations, we compare our LO
numerical results with the previous work in
Ref.\cite{B.A.Kniehl-lo}. We adopt the same input parameters as in
Ref.\cite{B.A.Kniehl-lo}, and reproduce the LO distribution of
$p_T^{J/\psi}$ which is in good agreement with that shown in
Fig.5(a) of Ref.\cite{B.A.Kniehl-lo}.

\par
In the following numerical calculations for the $pp\to J/\psi+Z^0+X$
process at the LHC, we take CTEQ6L1 PDFs with the one-loop running
$\alpha_s$ in the LO calculations and CTEQ6M PDFs with the two-loop
$\alpha_s$ in the NLO calculations \cite{CTEQ6}. For the QCD
parameters we take the number of active flavor as $n_f=3$, and input
$\Lambda_{{\rm QCD}}^{(3)} = 249~{\rm MeV}$ for the LO and
$\Lambda_{{\rm QCD}}^{(3)} = 389~{\rm MeV}$ for the NLO calculations
\cite{kniehl1}, respectively. The renormalization, factorization,
and NRQCD scales are taken as $\mu_r = \mu_f = m_T$ and
$\mu_{\Lambda} = m_c$, respectively, where $m_T = \sqrt{\Big(
p_T^{J/\psi} \Big)^2 + m_{J/\psi}^2}$ is the $J/\psi$ transverse
mass. The masses of the external particles and the fine structure
constant are taken as $m_Z =91.1876~{\rm GeV}$, $m_c = m_{J/\psi}/2
= 1.5~{\rm GeV}$ and $\alpha = 1/137.036$. The $\langle~ {\cal
O}^{J/\psi} [ ^3P_J^{(8)} ]~  \rangle$ $(J=0,1,2)$ LDMEs satisfy the
multiplicity relations
\begin{eqnarray}
\langle~ {\cal O}^{J/\psi} [ ^3P_J^{(8)} ]~
\rangle = (2J+1) \langle~ {\cal O}^{J/\psi} [
^3P_0^{(8)} ]~ \rangle, \nonumber
\end{eqnarray}
and the linear combination exists as
\begin{eqnarray}
M_r^{J/\psi} = \langle~ {\cal O}^{J/\psi} [ ^1S_0^{(8)} ]~ \rangle +
\frac{r}{m_c^2} \langle~ {\cal O}^{J/\psi} [ ^3P_0^{(8)} ]~ \rangle.
\nonumber
\end{eqnarray}
We take the LDMEs  $\langle~ {\cal O}^{J/\psi} [ ^3S_1^{(1)}]~
\rangle=1.3~{\rm GeV}^3$, $\langle~ {\cal O}^{J/\psi} [
^3S_1^{(8)}]~ \rangle=2.73 \times 10^{-3}~{\rm GeV}^3$,
$M_r^{J/\psi}=5.72 \times 10^{-3}~{\rm GeV}^3$ and $r = 3.54$ as the
input parameters, which were fitted to the Tevatron RUN-I data by
using the CTEQ4 PDFs and taking into account the dominant
higher-order effects due to the multiple-gluon radiation in the
inclusive $J/\psi$ hadroproduction \cite{Kniehl:1998qy}. Then we fix
the $\langle~ {\cal O}^{J/\psi} [ ^1S_0^{(8)} ]~ \rangle$ and
$\langle~ {\cal O}^{J/\psi} [ ^3P_0^{(8)} ]~ \rangle$ LDMEs by the
democratic choice \cite{Klasen:2004tz-a,Klasen:2004tz-b}
\begin{eqnarray}
\langle~ {\cal O}^{J/\psi} [ ^1S_0^{(8)} ]~
\rangle = \frac{r}{m_c^2} \langle~ {\cal O}^{J/\psi}
[ ^3P_0^{(8)} ]~ \rangle = \frac{1}{2}
M_r^{J/\psi}. \nonumber
\end{eqnarray}

\par
We have checked numerically the independence of the total NLO QCD
correction on the cutoffs $\delta_s$ and $\delta_c$ in the range
$1\times10^{-4} \leq \delta_s \leq 1\times 10^{-2}$ and
$\delta_c=\delta_s/50$. The independence exists in our numerical
results within the error tolerance. In further calculations, the two
phase space cutoffs are fixed as $\delta_s =1\times 10^{-3}$ and
$\delta_c = \delta_s/50$. Considering the validity of the NRQCD and
perturbation method, we restrict our results to the domain
$p_T^{J/\psi} > 3~{\rm GeV}$ and $|y_{J/\psi}| < 3$.
%%figure%%
\begin{figure}[htbp]
\vspace*{-0.3cm} \centering
\includegraphics[scale=1]{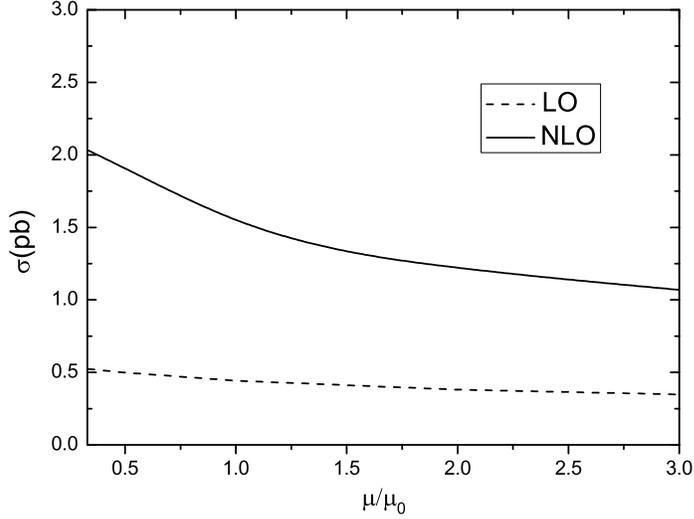}
\vspace*{-0.3cm} \centering \caption{\label{fig4} The dependence of
the LO and the NLO QCD corrected cross sections for the process
\ppJZ on the factorization scale and renormalization scale
($\mu$/$\mu_0$) at the LHC where we define $\mu=\mu_f=\mu_r$ and
$\mu_0 = m_T$. }
\end{figure}
%%%%figure%%%%

\par
The dependence of the cross section on the renormalization scale
$\mu_r$ and factorization scale $\mu_f$ induces uncertainty for
theoretical prediction. In Fig.\ref{fig4}, the $\mu$ dependence of
the LO and the NLO QCD corrected cross sections with the constraints
of $p_t^{J/\psi}
>3GeV$ and $|y_{J/\psi}| < 3$ for \ppJZ process, is shown with our default
choice $\mu$ = $\mu_r$ = $\mu_f$ and the definition of $\mu_0=m_T$.
There $\mu$ varies from $\mu_0/3$ to $3 \mu_0$. Not like the usual
expectation, Fig.\ref{fig4} shows that the NLO QCD correction can
not improve the LO scale independence for the \ppJZ process at the
LHC. Actually, the similar behavior appears also in the results at
the NLO in NRQCD (see Ref.\cite{jpsi:com-d}). The related
theoretical uncertainty amounts to  ${+13\atop-14}\%$ at the LO and
to ${+23\atop-21}\%$ at the NLO when $\mu$ goes from $\mu_0/2$ to $2
\mu_0$.
%%figure%%
\begin{figure}[htbp]
\vspace*{-0.3cm} \centering
\includegraphics[scale=1]{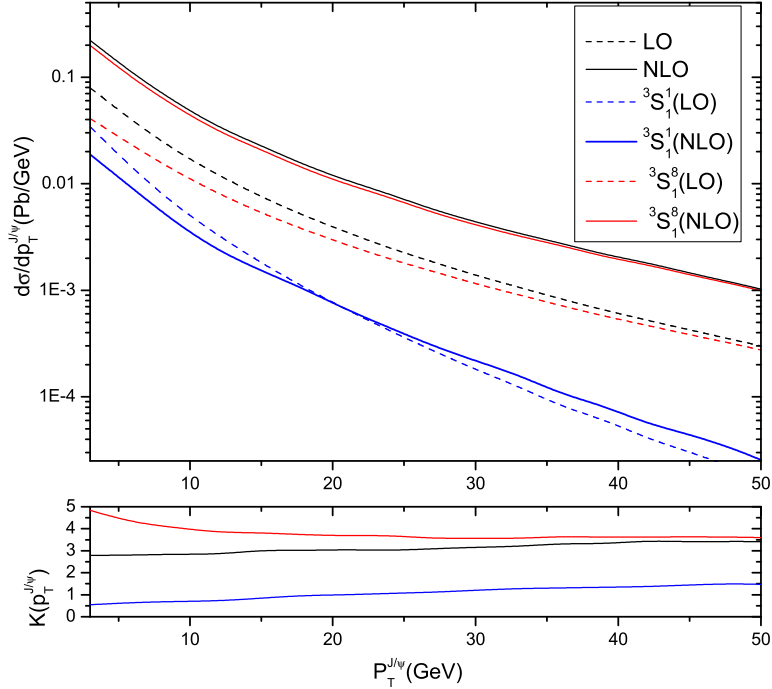}
\vspace*{-0.3cm} \centering \caption{\label{fig5} The LO and NLO QCD
corrected distributions of $p_T^{J/\psi}$ and the corresponding
K-factor for the \ppJZ process at the LHC. }
\end{figure}
%%%%figure%%%%

\par
In Fig.\ref{fig5}, we present the LO and NLO QCD corrected
distributions of $p_T^{J/\psi}$ and the corresponding K-factors for
the \ppJZ process at the LHC. For comparison, we also depict the LO
and NLO QCD contributions from the $c\bar{c}[^3S_1^{(1)}]$ and
$c\bar{c}[^3S_1^{(8)}]$ Fock states and their corresponding
K-factors in these figures separately. The upper figure shows that
the differential cross sections at the LO are significantly enhanced
by the NLO QCD corrections and both the LO and NLO QCD corrected
differential cross section curves drop rapidly with the increment of
$p_T^{J/\psi}$. We can read out from the figure that the LO
differential cross section of $p_T^{J/\psi}$ decreases from
$79.2~fb/GeV$ to $0.3~fb/GeV$ and the NLO QCD
corrected differential cross section of $p_T^{J/\psi}$ decreases
from $221.1~fb/GeV$ to $1~fb/GeV$ as $p_T^{J/\psi}$ goes up from $3$
to $50~GeV$. The corresponding
$K$-factor, defined as $K =
\frac{d\sigma^{NLO}}{dp_T^{J/\psi}}/\frac{d\sigma^{LO}}{dp_T^{J/\psi}}$,
varies in the range of $[2.79,~3.42]$, and reach its maximum when
$p_T^{J/\psi} = 50~GeV$.

\par
From Fig.\ref{fig5}, we find that the contributions to the
differential cross section from the $c\bar{c}[^3S_1^{(1)}]$ and
$c\bar{c}[^3S_1^{(8)}]$ Fock states are comparable at lower
$p_T^{J/\psi}$. As the $p_T^{J/\psi}$ increases, the contribution of
$c\bar{c}[^3S_1^{(1)}]$ decreases much faster than
$c\bar{c}[^3S_1^{(8)}]$. In all the plotted $p_T^{J/\psi}$ range,
the CO contributions to the differential cross section of the
process of \ppJZ are dominant, and that will be beneficial to our
study of CO mechanism. For the contribution from the
 $pp \to c\bar{c}[^3S_1^{(1)}]+Z^0+X$ process,
the $K$-factor of differential cross section increases from $0.55$ to $1.50$
as $p_T^{J/\psi}$ increases from $3~GeV$ to $50~GeV$. As for the $pp \to
c\bar{c}[^3S_1^{(8)}]+Z^0+X$ process, the $K$-factor is larger than
that for the $pp \to c\bar{c}[^3S_1^{(1)}]+Z^0+X$ process, which
decreases from $4.9$ to $3.6$ as $p_T^{J/\psi}$ increases from
$3~GeV$ to $50~GeV$.

\vskip 5mm
\section{Summary}
\par
The $J/\psi+Z^0$ associated production is an important process in
investigating the production mechanism of $J/\psi$. In this paper we
investigate the NLO QCD corrections to the $J/\psi +Z^0$ production
at the LHC, and present the numerical predictions of the
$p_T^{J/\psi}$ distribution of the $p_T^{J/\psi}$ up to the QCD NLO.
We find that the LO differential cross section for the process $pp
\to J/\psi+Z^0+X$ is heavily enhanced by the NLO QCD corrections,
and the K-factor can reach the value of $3.42$ in the large
$p_T^{J/\psi}$ region. The NLO QCD corrected differential cross
section can reach $221.1~fb/GeV$ in the vicinity of $p_T^{J/\psi}
\sim 3~GeV$. The process of $J/\psi + Z$ associated production is
dominantly contributed by the CO mechanism in the large
$p_T^{J/\psi}$ range. We conclude that the LHC has the potential to
study the $J/\psi+Z^0$ production process, by which we can
investigate the production mechanism of heavy quarkonium and extract
the universal NRQCD matrix elements.

\vskip 5mm
\par
\noindent{\large\bf Acknowledgments:} This work was supported in
part by the National Natural Science Foundation of
China(No.10875112, No.11075150, No.11005101), the Specialized
Research Fund for the Doctoral Program of Higher
Education(No.20093402110030), and the 211 Project of Anhui
University.

\end{document}